\documentstyle{article}
\begin{document}
\newcommand{\beq}{\begin{equation}}
\newcommand{\eeq}{\end{equation}}
\newcommand{\beqn}{\begin{eqnarray}}
\newcommand{\eeqn}{\end{eqnarray}}
\newcommand{\dpf}{\displaystyle\frac}
\newcommand{\no}{\nonumber}
\newcommand{\ep}{\epsilon}
\begin{center}
{\Large Entropy of extreme three-dimensional charged black holes}
\end{center}
\vspace{1ex}
\centerline{\large Bin
Wang$^{a,b,}$\footnote[1]{e-mail:binwang@fma.if.usp.br},
\ Elcio Abdalla$^{a,}$\footnote[2]{e-mail:eabdalla@fma.if.usp.br}}
\begin{center}
{$^{a}$ Instituto De Fisica, Universidade De Sao Paulo, C.P.66.318, CEP
05315-970, Sao Paulo, Brazil \\
$^{b}$ Department of Physics, Shanghai Teachers' University, P. R. China}
\end{center}
\vspace{6ex}
\begin{abstract}
It is shown that three-dimensional charged black holes can approach the
extreme state at nonzero temperature. Unlike even dimensional cases, the
entropy for the extreme three-dimensional charged black hole is uniquely
described by the Bekenstein-Hawking formula, regardless of different
treatments of preparing the extreme black hole, namely, Hawking's
treatment and Zaslavskii's treatment. 
\end{abstract}
\vspace{6ex} \hspace*{0mm} PACS number(s): 04.70.Dy, 04.20.Gz, 04.62.+v.
\vfill
\newpage
There has been recently a great deal of interest in the  study of the
extreme
black hole (EBH) entropy. The interest was first heated up by the findings
that the four-dimensional (4D) Reissner-Nordstr$\ddot{o}$m (RN) EBH and
the non-extreme black hole (NEBH) are different objects due to their
drastically different topological properties and RN EBH has zero entropy
regardless of its nonzero horizon area [1,2]. However, using the  grand
canonical ensemble, Zaslavskii argued that a 4D RN black hole in a finite
size cavity can approach the extreme state as closely as one likes and the
Bekenstein-Hawking formula is still expected to hold for RN EBH [3]. The
geometrical and topological properties were also claimed of nonextreme
sectors [4,5]. Support for this view is also provided by state-counting
calculations of certain extreme and near-extreme black holes in string
theory, see [6] for a review. These different results indicate that EBHs
have a special and controversial role in black hole thermodynamics and
topologies. 

Comparing [1,2] and [3-5], it seems that the clash comes from two
different treatments: one refers to Hawking's treatment by starting with
the original EBH [1,2] and the other Zaslavskii's treatment by first
taking
the boundary limit and then the extreme limit to get the EBH from its
nonextreme counterpart [3-5]. Recently by using these two treatments, the
geometry and intrinsic thermodynamics have been investigated in detail for
a wide class of EBHs including 4D and two-dimensional (2D) cases
[7-10]. It was found that these different treatments lead to two different
topological objects represented by different Euler characteristics and
show drastically different intrinsic thermodynamical properties both
classically and quantum-mechanically. Based upon these results it was
suggested that there maybe two kinds of EBHs in the nature: the first kind
suggested by Hawking et al with the extreme topology and zero entropy,
which can only be formed by pair creation in the early universe; on the
other hand,
the second kind, suggested by Zaslavskii, has the topology of
the nonextreme sector
and the entropy is still described by the Bekenstein-Hawking formula,
which can be developed from its nonextreme counterpart through second
order phase transition [11-13]. This speculation has been further
confirmed recently in a Hamiltonian framework [14] and the grand canonical
ensemble [15] as well as canonical ensemble [16] formulation for RN
anti-de Sitter black hole.

All these results available for EBHs' entropy are limited to even
dimensions. Whether these results can be extended to 
odd dimensions is unclear. This paper evolves from an attempt
to study this problem by using (2+1)-dimensional (3D) charged black hole
as an example.

The metric of the 3D charged black hole reads [17]
\beq                
{\rm d}s^2=-N^2{\rm d}t^2+N^{-2}{\rm d}r^2+r^2{\rm d}\phi^2
\eeq
where
\beq 
N^2=-M+\dpf{r^2}{l^2}-\dpf{\ep^2}{2}\ln\dpf{r}{r_0}
\eeq
with $-\infty<t<+\infty, 0<r<\infty$ and $0\leq\phi\leq 2\pi$, $M$ and
$\ep$
in the above metric associated respectively with the mass and the charge
of the black hole, $-l^{-2}$
is the negative cosmological constant and $r_0$ is a constant. When
$r>r_0$, the 3D charged black hole is described by  
the Penrose diagram as usual[18]. The electric potential of the charge is
\beq           
A_0(r)=-\ep\ln\dpf{r}{r_0}.
\eeq
This black hole has two, one, or no horizons, depending on whether [19]
\beq 
M-(\dpf{\ep^2}{4}-\dpf{\ep^2}{4}\ln\dpf{\ep^2 l^2}{4r_0^2})
\eeq 
is greater than, equal to or less than zero, respectively.

Now we can directly make use of the approach of [20] to study the black
hole thermodynamics in a grand canonical ensemble where we consider the
black hole in a cavity with radius $r_B$. The temperature on the
boundary of the cavity is $T_W=T_H/N(r_B)$, where $T_H = k/2\pi$
is the Hawking temperature and $k$ is the surface gravity.

For our metric (1), the local temperature has the form
\beqn
T_W & = & \dpf{T_H}{\sqrt{-M+r_B^2/l^2 - \ep^2 /2 \ln(r_B/r_0)}} \\
T_H & = & \dpf{2r_+/l^2 - \ep^2 /2r_+}{4\pi}
\eeqn
When a black hole approaches the extreme state
$(M=\dpf{\ep^2}{4}-\dpf{\ep^2}{4}\ln\dpf{\ep^2 l^2}{4r_0^2},
\ep^2=\dpf{4r_+ ^2}{l^2})$, according to (6) $T_H\rightarrow 0$. The
simplest choice is to take the limit $T_W\rightarrow 0$. One might refer
to the third law of thermodynamics to argue that the EBH cannot be
achieved because the absolute zero temperature is unachievable.

However, it is interesting to point out that although $T_H\rightarrow
0$, the
square root in (5) tends to zero as well if we take $r_+\rightarrow r_B$,
thus the extreme state with nonzero local temperature does exist.
Indeed, taking $r_+$ and $r_-$ as corresponding to event horizon
and Cauchy horizons, we have
\beq
\ep^2=\dpf{2(r_+^2-r_-^2)}{l^2\ln (r_+/r_-)},
\eeq
and we can readily see that although $T_H$ has a simple zero in
$r_+\rightarrow r_-$, the expression in the square-root in the denominator
of $T_W$ has a double-zero, i.e.,
\beq
-M+\dpf{r_B^2}{l^2}-\dpf{\ep^2}{2}\ln\dpf{r_B}{r_0}=
\dpf{r_+^2-r_-^2}{l^2}(1-\dpf{\ln r_+ -\ln r_-}{\ln r_+ -\ln r_-}),
\eeq
therefore $T_W$ tends to a constant value in the EBH case.
 Recall
that in the grand canonical ensemble, only the temperature on the boundary
has physical meaning, whereas $T_H$ can always be rescaled without
changing observable quantities [21]. Therefore analogous to the 4D RN case
[3], there exists a well defined extreme state of the 3D charged black
hole in the grand canonical ensemble and no contradiction with the third
law arise.

Now it is of interest to investigate whether two different treatments
applied in even dimensions will lead to similar different entropy results
for 3D charged EBH. The action for the Euclidean version of the 3D charged
black hole on a 3D manifold $M$ with a boundary is given by

\beq
I=-\dpf{1}{2\pi}\int_M d^3 x\sqrt{-g}(R+2/l^2+1/4F_{\mu\nu}F^{\mu\nu})
+\dpf{1}{\pi}\int_{\partial M} d^2 x\sqrt{-\gamma} (K-K_0)
\eeq
Here $\gamma$ is the induced metric on the boundary $\partial M$ and $K$
is the extrinsic curvature of the boundary. $K_0$ is a constant
independent on the metric of 3D spacetime and we choose it to be zero to
normalize the thermodynamic energy in a flat spacetime.

Introducing the Gaussian normal coordinates
near every point on the surface of the cavity, the timelike coordinate of
this system is
the proper
time $\tau$ for an observer on the surface and the coordinates on
the surface
are ($\tau, \phi$).
Defining $\vec{N}$ as the unit spacelike vector orthogonal to the surface 
and
$\vec{U}$ the velocity
of a mass element of this surface, the orthogonal condition becomes
\beq        
\vec{N}\cdot\vec{U}=0
\eeq
The velocity is $\vec{U}=\dot{t}\partial_t+\dot{r}\partial_r$
where
the overdot denotes differentiation with respect to $\tau$. We obtain
$\vec{N}=(\vert g_{tt}\vert)^{-1}\dot{r}\partial_t+\vert
g_{tt}\vert\dot{t}\partial_r$
from Eq(10). The normalization conditions are $\vec{N}\cdot\vec{N}=1,
\vec{U}\cdot\vec{U}=-1$.
The extrinsic curvatures relative to the Gaussian normal coordinates are
\beqn         
K_{\tau\tau}=N_{\tau;\tau} & = & U^{\mu}U^{\nu}N_{\mu\nu}  \\
              K_{\phi\phi} & = & N_{\phi;\phi}
\eeqn

The action for the black hole with the cavity at $r=r_B$ is 
\beq
\dpf{\beta}{N_{r_B}}[\dpf{4}{l^2}(r_B^2-r_+^2)-\dpf{\ep^2}{4}\ln\dpf{r_B}{r_+}]
+2\beta[\dpf{r_B}{2N_{r_B}}(\dpf{dN^2}{dr})_{r_B}+N_{r_B}],
\eeq
where the relation $\beta=T^{-1}_W=\int_0^{2\pi} N(r_B) d\tau$ has been
used.

The free energy is given by the expression
\beq
F=\dpf{I}{\beta}=\dpf{1}{N_{r_B}}[\dpf{4}{l^2}(r_B^2-r_+^2)-\dpf{\ep^2}{4}\ln
\dpf{r_B}{r_+}]+2\dpf{r_B}{2N_{r_B}}(\dpf{dN^2}{dr})_{r_B}+2N_{r_B},
\eeq
while the entropy can be calculated by means of the formula
\beq
S=-(\dpf{\partial F}{\partial T_W})_D =-(\dpf{\partial F}{\partial r_+})_D
(\dpf{d T_W}{d r_+})^{-1}_D.
\eeq
We have, 
\beqn
S & = & - 4\pi\dpf{(-8r_+/l^2 +\ep^2/4r_+)N_{r_B}+\dpf{d N_{r_B}}{d
r_+}[4(r_B^2-r_+^2)/l^2-\ep^2/4\ln(r_B/r_+)]}
{\dpf{d N_{r_B}}{d r_+}(\dpf{d N^2}{d r})_{r_+} - N_{r_B}\dpf{d}{d
r_+}(\dpf{d N^2}{d r})_{r_+}} \no \\ 
  &   & -8\pi N^2_{r_B}\dpf{-\dpf{r_B}{2N^2_{r_B}}\dpf{d N_{r_B}}{d r_+}
(\dpf{d N^2}{d r})_{r_B}+\dpf{r_B}{2N_{r_B}}\dpf{d}{d r_+}(\dpf{d N^2}{d
r})_{r_B}+\dpf{d N_{r_B}}{d r_+}}{\dpf{d N_{r_B}}{d
r_+}(\dpf{d N^2}{d r})_{r_+} - N_{r_B}\dpf{d}{d r_+}(\dpf{d N^2}{d
r})_{r_+}}.
\eeqn
Taking the boundary limit $r_+\rightarrow r_B, (N_{r_B}\rightarrow 0)$, we
find
\beq
S=4\pi r_+
\eeq
This is just the entropy for the 3D charged NEBH [17]. We note that the
first term in (16) does not contribute to the entropy, which is similar to
the even dimensional cases, where the  entropy result is only attributed
to the surface term of the Euclidean action.

We are now in position to extend the above calculations to EBH. We are
facing two limits, namely, the boundary limit $r_+\rightarrow r_B$ and the
extreme limit $M=\dpf{\ep^2}{4}-\dpf{\ep^2}{4}\ln\dpf{\ep^2 l^2}{4r_0^2},
\ep^2=4r_+ ^2/l^2$. We follow two different treatments while taking these two
limits: (A) first take the boundary limit and then the extreme limit,
which corresponds to the treatment adopted in [3-5]; and (B) first take
the extreme limit and then the boundary limit, which corresponds to
starting with the original EBH in [1,2]. From Eq.(14), it is easy to find
that both the first term and the third term in the free energy will vanish
either in treatment (A) or (B) due to the limit $r_+\rightarrow r_B$.
Therefore only the second term of the free energy has the contribution to
the entropy in these two treatments. Using (15), we have
\beqn
S(A) & = & [\dpf{4\pi r_B \dpf{d N_{r_B}}{d r_+}(\dpf{d N^2}{d
r})_{r_B}}{\dpf{d
N_{r_B}}{d r_+}(\dpf{d N^2}{d r})_{r_+}-N_{r_B}\dpf{d}{d r_+}(\dpf{d
N^2}{d r})_{r_+}}]_{r_+\rightarrow r_B}\vert_{extr} =4\pi r_+ \\
S(B) & = & [\dpf{4\pi r_B \dpf{d N_{r_B}}{d r_+}(\dpf{d N^2}{d
r})_{r_B}}{\dpf{d
N_{r_B}}{d r_+}(\dpf{d N^2}{d r})_{r_+}-N_{r_B}\dpf{d}{d r_+}(\dpf{d      
N^2}{d r})_{r_+}}]_{extr}\vert_{r_+\rightarrow r_B} \\ \no
  &    & =lim_{r_+\rightarrow r_B} 4\pi
r_B\dpf{r_+}{r_B}\dpf{\ln\dpf{r_B^2}{r_+^2} (\dpf{r_B^2}{l^2
r_+^2}-\dpf{1}{l^2})}{\ln\dpf{r_B^2}{r_+^2}(1/l^2-1/l^2)-2(\dpf{r_B^2}{r_+^2}-1)(1/l^2
-1/l^2)} =4\pi r_+
\eeqn
These two different ways of taking the limits lead to the same
entropy for 3D charged EBH, and entropy never vanishes. This result
can also be extended to 3D rotating black hole.

Thus we have shown that in the grand canonical ensemble, the 3D charged
black hole can approach the extreme state at nonzero temperature. Unlike
even dimensional cases, the entropy of the 3D charged EBH is uniquely
described by the Bekenstein-Hawking formula regardless of the different
ways of taking the limits. 

As a matter of fact, in even dimensions it is usual to classify the
topology of the manifold in terms of the Chern class, in dimensions
multiple of four also by the Pontryagin number. The problem of the black
hole entropy has been generally related to the Euler characteristic of the
manifold, rather useful in the context of general relativity due to the
Gauss-Bonnet theorem. Thus, for even dimensional NEBHs and EBHs, there are
direct relations between the black hole entropy and the topological properties
represented by Euler characteristics obtained from Gauss-Bonnet theorem
[22,23,24,25,2,8-11]. In odd dimensional space-time the situation is much
more difficult, since most of the tradicional topological invariants do
not exist, in spite of the fact that the topology may be far from trivial.
In three dimensions, in particular, the Gauss-Bonnet theorem
does not exist, and the relations between the entropies
and topologies either in NEBH or EBH valid in even dimensions are not
appropriate. As far as the traditional invariants are concerned, our
result is compatible with having the same topology for both, the extreme 
and non-extreme black holes in three dimensions, thus we find no 
contradiction when the extreme limit is taken. However, for more general
configurations we certainly need a finer analysis. We can say that
the relation between the entropy result and their topological
properties in odd dimensions is still unclear and needs further study.

ACKNOWLEDGEMENTS: This work was partically supported
by Fundac\~{a}o de Amparo \`{a} Pesquisa do Estado de
S\~{a}o Paulo (FAPESP) and Conselho Nacional de Desenvolvimento 
Cient\'{\i}fico e Tecnol\'{o}gico (CNPQ).  B. Wang would also
like to acknowledge the support given by Shanghai Science and Technology
Commission. We would like to thank Dr. A. Saa for discussions.

\end{document}